\documentclass[sigconf]{acmart}

\AtBeginDocument{%
  }

\usepackage[table]{xcolor}
\usepackage{booktabs}
\usepackage{geometry}
\usepackage{soul}

\usepackage{pifont}

\usepackage{pgfplots}
\pgfplotsset{compat=1.18}
\usepgfplotslibrary{groupplots}

\usepackage{todonotes}
\presetkeys%
    {todonotes}%
    {inline,backgroundcolor=yellow}{}

\usepackage{graphicx}
\usepackage{epstopdf}

\usepackage{tabularx}
\usepackage{booktabs}
\usepackage{array}

\usepackage{mdframed}
\newmdenv[
  backgroundcolor=gray!15, 
  linecolor=gray,          
  linewidth=3pt,            
  leftline=true,            
  topline=false,
  bottomline=false,
  rightline=false,
  innertopmargin=1ex,
  innerleftmargin=1ex,
]{myprompt}

\usetikzlibrary{patterns}

\usepackage{wrapfig}

\begin{document}

\renewcommand{\arraystretch}{1.4} 
\setlength{\tabcolsep}{10pt}      
\rowcolors{2}{gray!10}{white}     

\newcommand{\cmark}{\ding{51}}
\newcommand{\xmark}{\ding{55}}

 \newcommand{\faketodo}[1]{\textcolor{red}{#1}}
\newcommand{\faketodobrackets}[1]{\textcolor{red}{[#1]}}
\newcommand{\faketodoitalic}[1]{\textcolor{red}{\textit{#1}}}

\title{Harnessing the Power of Large Language Models for Software Testing Education: A Focus on ISTQB Syllabus}

\author{Tuan-Phong Ngo}
\email{phong.ngo@rmit.edu.vn}
\orcid{0000-0003-4993-0092}
\affiliation{%
  \institution{RMIT University Vietnam}
  \city{Hanoi}
  \country{Vietnam}
}

\author{Bao-Ngoc Duong}
\email{s3425449@rmit.edu.vn}
\affiliation{%
  \institution{RMIT University Vietnam}
  \city{Hanoi}
  \country{Vietnam}
}

\author{Tuan-Anh Hoang}
\email{anh.hoang62@rmit.edu.vn}
\orcid{0000-0003-3892-4762}
\affiliation{%
  \institution{RMIT University Vietnam}
  \city{Hanoi}
  \country{Vietnam}
}

\author{Joshua Dwight}
\email{joshua.dwight@rmit.edu.vn}
\orcid{0000-0002-6247-2703}
\affiliation{%
  \institution{RMIT University Vietnam}
  \city{Hanoi}
  \country{Vietnam}
}

\author{Ushik Shrestha Khwakhali}
\email{ushik.shrestha@rmit.edu.vn}
\orcid{0000-0002-8453-7926}
\affiliation{%
  \institution{RMIT University Vietnam}
  \city{Hanoi}
  \country{Vietnam}
}


\begin{abstract}

Software testing is a critical component in the software engineering field and important for software engineering education. Thus, it is vital for academia to continuously improve and update educational methods to reflect the current state of the field. The International Software Testing Qualifications Board (ISTQB) certification framework is globally recognized and widely adopted in industry and academia. However, ISTQB-based learning has been rarely applied with recent generative artificial intelligence advances. Despite the  growing capabilities of large language models (LLMs), ISTQB-based learning and instruction with LLMs have not been thoroughly explored.
 
This paper explores and evaluates how LLMs can complement the ISTQB framework for higher education. The findings present four key contributions: ($i$) the creation of a comprehensive ISTQB-aligned dataset spanning over a decade, consisting of 28 sample exams and 1,145 questions; ($ii$) the development of a domain-optimized prompt that enhances LLM precision and explanation quality on ISTQB tasks; ($iii$) a systematic evaluation of state-of-the-art LLMs on this dataset; and ($iv$) actionable insights and recommendations for integrating LLMs into software testing education. These findings highlight the promise of LLMs in supporting ISTQB certification preparation and offer a foundation for their broader use in software engineering at higher education.

\end{abstract}




\maketitle

\section{Introduction}
\label{intro:sect}

Software testing is a fundamental practice in software engineering, essential for verifying that software systems fulfill specified requirements related to quality, reliability, and user satisfaction~\cite{Myers2012}. It plays a critical role in assessing software quality and mitigating risks of operational failures. However, the efficacy of software quality significantly declines when testing is not conducted systematically and rigorously.

Despite its importance, comprehensive testing poses considerable challenges. It is prone to errors and often perceived as tedious by software developers, leading to its deprioritization in routine workflows~\cite{Clegg2017}. Moreover, many recent graduates entering the software industry have limited practical experience in software testing~\cite{Zivkovic2021}, highlighting the imperative to strengthen software testing education at the university level.

This need is accentuated by the significant growth of the global software market, driven by increasing digital transformation across industries. The market size is projected to expand at a compound annual growth rate (CAGR) of 3.98\% from 2025 to 2030, reaching USD 902.74 billion~\cite{statistaSoftwareWorldwide2025}. Similarly, the global software testing sector is experiencing rapid growth, with its market value expected to rise from USD 117.01 billion in 2025 to over USD 436.62 billion by 2033, at a robust CAGR of 17.9\%~\cite{SoftwareTestingReport2025}.

The International Software Testing Qualifications Board (ISTQB), a globally recognized certification authority, recently surpassed one million certifications awarded across more than 130 countries~\cite{istqb}. This milestone underscores the rising demand for skilled software testing professionals and reinforces the necessity of prioritizing software testing within software engineering curricula.

In this evolving context, Large Language Models (LLMs) have emerged as indispensable tools for enhancing software testing processes. Their applications include generating unit tests, designing test cases, debugging, bug fixing, fault localization, and automated program repair~\cite{LI2025103942, Wang2024}. Leading models such as ChatGPT, Gemini, Claude, Llama, and Mistral demonstrate superior speed and accuracy over alternative approaches~\cite{statistaLLM2025}.

The global LLM market is poised for rapid expansion, forecasted to grow from USD 1.59 billion in 2023 to USD 259.8 billion by 2030, with a CAGR of 79.8\%. By 2025, approximately 750 million applications are expected to utilize LLMs, automating up to 50\% of digital tasks~\cite{springsappsLLM2024}. These trends highlight the vital role of LLMs in supporting ISTQB certification preparation and advocate for their broader integration into higher education software engineering programs.

The key contributions of this research are as follows:
\begin{enumerate}
\item Compilation of a comprehensive ISTQB-aligned dataset, comprising 28 sample exams with 1,145 questions.
\item Design and implementation of a domain-specific prompt that significantly enhances LLM accuracy and explanation quality on ISTQB tasks.
\item Systematic evaluation of state-of-the-art LLMs on this dataset to benchmark their performance.
\item Provision of actionable insights and recommendations to integrate LLMs into software testing education.
\end{enumerate}

\section{Background}
\label{background:sect}

\subsection{The ISTQB Certification Framework}
\label{istqb:subsect}
The \emph{International Software Testing Qualifications Board} (ISTQB) provides a globally recognized certification framework for software testing professionals~\cite{istqb}. Its syllabus covers foundational to advanced testing concepts, offering a structured pathway for industry-relevant knowledge~\cite{Dorothy2019,Catherine2019}. 

The certification is structured into \emph{three  levels}: Foundation, focusing on core principles and terminology; Advanced, targeting analytical and managerial skills for roles like test analyst and manager; and Expert, emphasizing strategic leadership and domain-specific depth.

ISTQB certifications are organized into  \emph{three streams}: Core, covering general testing principles; Agile, addressing practices in Agile and DevOps environments; and Specialist, targeting areas such as test automation and security.

Each level and stream of the exam assesses learners at varying \emph{cognitive levels}, aligned with Bloom’s Taxonomy~\cite{Krathwohl01112002}. This approach provides progressively higher cognitive demands, ensuring a thorough evaluation of both theoretical understanding and practical skills (more details in Section~\ref{istqb_benchmark:sect}.)

\subsection{Integrating the ISTQB Syllabus into Undergraduate Curriculum}
\label{software_testing_istqb:subsect}

Incorporating the ISTQB syllabus into undergraduate education effectively bridges the gap between academic theory and industry practice, thereby enhancing graduate employability and aligning student skills with established professional standards. Several academic programs have begun integrating the ISTQB syllabus into their software testing courses, leveraging its comprehensive framework to increase curriculum rigor and relevance~\cite{Jokisch2025, Garousi2024}. This approach benefits students by providing well-defined learning objectives and preparing them for widely recognized certification exams.

Nevertheless, educators encounter \emph{challenges} in adapting the syllabus to diverse student backgrounds, balancing theoretical instruction with practical application, and managing limited resources~\cite{10132212}. \emph{Opportunities} exist to address these challenges through innovative pedagogical approaches, such as active learning techniques and the incorporation of technology-enhanced tools, which can improve student engagement and learning outcomes~\cite{10132212}. This study focused on integrating the Foundation and Advanced levels of the ISTQB framework into the undergraduate curriculum.

\subsection{The Emergence of Large Language Models in Software Testing Education}
\label{llms:subsect}
\emph{Large Language Models} (LLMs), such as GPT-based architectures, are revolutionizing educational practices by enabling personalized, interactive, and scalable learning experiences.
In software testing, LLMs have demonstrated capabilities in automating test case generation, producing documentation, and assisting in code reviews, thereby streamlining testing workflows~\cite{LI2025103942, Wang2024}.
Beyond these tasks, LLMs offer substantial potential in education by functioning as intelligent assistant, delivering real-time feedback, and supporting adaptive learning experiences~\cite{KASNECI2023102274}.
%


\subsection{Gaps in Literature}
\label{gaps:subsect}
Despite growing interest in LLM applications for software testing, \emph{significant gaps} remain in understanding their educational integration, particularly aligned with established standards like the ISTQB syllabus.

Existing research primarily evaluates the technical capabilities of LLMs in automated test generation and program repair~\cite{Ebert2023,chen2021codex}, but insufficient attention has been given to the practical challenges faced by educators and learners in software testing contexts. These challenges include the complexity of prompt engineering, difficulties in interpreting LLM outputs, and the need to tailor AI tools to accommodate diverse learner proficiencies.

Moreover, research exploring how LLMs can be systematically embedded within ISTQB-based curricula to support both students and instructors is limited. For students, LLMs hold promise as virtual tutors that facilitate mastery through interactive dialogue and tailored explanations. For instructors, LLMs can aid in identifying difficult syllabus topics, developing instructional content, and devising innovative teaching strategies. Addressing these gaps is essential to fully realize the pedagogical benefits of LLMs in software testing education and to foster standards-aligned learning environments~\cite{KASNECI2023102274}.

\section{Study Design}
\label{study_design:sect}

\subsection{Research Questions}
\label{research_questions:sect}
We study \emph{three} key research questions to explore the practical integration of LLMs to ISTQB-based software testing curricula:
\begin{itemize}
\item \textbf{RQ1} - To what extent can LLMs accurately address questions derived from the ISTQB exams?

\item \textbf{RQ2} - How can prompt engineering  be optimized to enhance the accuracy and pedagogical value of LLM-generated responses to ISTQB-related queries?


\item \textbf{RQ3} - How can LLMs be utilized to improve student engagement and learning outcomes?
\end{itemize}

\subsection{Open-source ISTQB Benchmark}
\label{istqb_benchmark:sect}

The ISTQB benchmark dataset presented in Table~\ref{tab:istqb-dataset}, compiled from the ISTQB organization, encompasses exam samples across various streams and levels, as detailed in Section~\ref{istqb:subsect}. 
Exceptions include CTAL-ATT v1.1 and CTAL-TM v2.0, which lacks sufficient question samples to form complete exams, and the Expert level, where essay questions necessitate a complex evaluation mechanism.

\begin{table}[tb]
\caption{ISTQB Dataset: St = Stream (C: Core, A: Agile, S: Specialist); L = Level (Fo: Foundation, Ad: Advanced, NA: Not Applicable); K = Cognitive Levels; \#Q = Number of questions per exam; \#P = Total possible points; \#PS = Passing score; \#QS = Total question samples.}
\label{tab:istqb-dataset}
\small
\centering
\setlength{\tabcolsep}{6pt}
\begin{tabular}{lll|c|ccc|c}
\rowcolor{gray!25}
\toprule
\textbf{St} & \textbf{L} & \textbf{Acronym}      & \textbf{K}  & \textbf{\#Q} & \textbf{\#P}             & \textbf{\#PS} & \textbf{\#QS}\\
\midrule
C      & Fo       & CTFL v4.0.1           &1-3                 & 40                & 40                       & 26                       & 160\\
C      & Fo       & CTFL v3.1             &1-3                 & 40                & 40                       & 26                       & 120 \\

C      & Ad       & CTAL-TM v3.0          &2-4                 & 50                & 88                       & 58                       & 50\\
C      & Ad       & CTAL-TA v4.0          &2-4                 & 45                & 78                       & 51                       & 45\\
C      & Ad       & CTAL-TA v3.1          &2-4                 & 40                & 80                       & 52                       & 40\\
C      & Ad       & CTAL-TAE v2.0         &2-4                 & 40                & 66                       & 43                       & 40\\
C      & Ad       & CTAL-TTA v4.0         &2-4                 & 45                & 78                       & 51                       & 45\\
A      & Fo       & CTFL-AT v1.0          &1-3                 & 40                & 40                       & 26                       & 40\\
S      & NA       & CT-AcT v1.0           &1-3                 & 40                & 40                       & 26                       & 40\\
S      & NA       & CT-AI v1.0            &1-4                 & 40                & 47                       & 31                       & 40\\
S      & NA       & CT-ATLaS v2.0         &2-4                 & 40                & 71                       & 47                       & 40\\
S      & NA       & CT-AuT v2.1           &1-3                 & 40                & 40                       & 26                       & 40\\
S      & NA       & CT-AuT v2.0           &1-3                 & 40                & 40                       & 26                       & 40\\
S      & NA       & CT-GaMe v1.0.1        &1-3                 & 40                & 40                       & 26                       & 40\\
S      & NA       & CT-GT v1.0.1            &1-2                 & 40                & 40                       & 26                       & 40\\
S      & NA       & CT-MAT v1.3           &1-3                 & 40                & 40                       & 26                       & 40\\
S      & NA       & CT-MBT v1.1           &1-3                 & 40                & 40                       & 26                       & 40\\
S      & NA       & CT-PT v1.0            &1-4                 & 40                & 40                       & 26                       & 40\\
S      & NA       & CT-SEC v1.1           &2-4                 & 45                & 80                       & 52                       & 45\\
S      & NA       & CT-STE v1.0.1         &2-4                 & 40                & 43                       & 28                       & 40\\
S      & NA       & CT-TAE v1.0           &2-4                 & 40                & 75                       & 49                       & 40\\
S      & NA       & CT-TAS v1.0           &2-3                 & 40                & 49                       & 32                       & 40\\
S      & NA       & CT-UT v1.2            &1-4                 & 40                & 40                       & 26                       & 40\\

\bottomrule
\end{tabular}
\end{table}

Spanning from 2015 to 2025, this collection includes 28 sample exams with a total of 1,145 multiple-choice questions, reflecting the evolution of software testing practices and knowledge requirements over a decade. Each exam is characterized by its total number of questions, possible points, and the passing score necessary for successful completion.
Accompanying the exams are official solution sets that provide both the correct answers and corresponding explanations.

To maintain the dataset's challenge for LLMs, we provide an exam \emph{generator} that automatically selects questions in accordance with the ISTQB exam structure and randomly shuffles answer options (see Figure~\ref{eval_framework:fig}). Experiments in Section~\ref{rq1_results:sect} demonstrate that this ISTQB question collection and the accompanying generator are valuable for evaluating and understanding the performance of LLMs on ISTQB certifications across various areas of expertise. This dataset will be released as an open-source resource in conjunction with this paper.

\subsection{Prompt Engineering Design}
\label{prompt_engineering:sect}
We developed \emph{two} levels of prompt complexity to assess the impact of prompt engineering on LLM performance in ISTQB exam contexts. The \textit{\textbf{basic level}} simulates a typical student interaction by directly posing original ISTQB questions to the LLMs, without contextual framing, role assignment, or structured output guidance. The \textit{\textbf{advanced level}} enhances response precision, comprehensiveness, and domain fidelity by utilizing sophisticated prompt engineering. It incorporates the role of an expert, a multi-part response format, and explicit alignment with the ISTQB syllabus. 

These two prompt designs tailored for students and educators avoid the complexity of techniques like few-shot prompting~\cite{NEURIPS2020_1457c0d6} or retrieval-augmented generation (RAG)~\cite{NEURIPS2020_6b493230}. The self-contained design allows these prompts to be easily deployed via standard chat interfaces, enhancing accessibility in educational settings. While advanced methods such as few-shot prompting or RAG can improve response accuracy and explanation relevance by integrating external knowledge or curated examples, they require sophisticated architectures and a deeper understanding of LLM internals, making them less suitable for typical educational use.
Due to space limitation, more detailed prompt information is provided in the Appendix.

\subsection{LLM Models}
\label{llm_selection:sect}

\setlength{\columnsep}{10pt}
\begin{wrapfigure}[10]{r}{0.27\textwidth}
\vspace{-13pt}
\centering
\setlength{\tabcolsep}{4pt}
\begin{tabular}{ll}
\rowcolor{gray!25}
\toprule
\textbf{Non-reasoning} & \textbf{Reasoning}\\
\midrule
\emph{\textsf{gpt-3.5-turbo}} & \textsf{o4-mini}  \\
\textsf{gpt-4o} & \textsf{o3}  \\
\bottomrule
\end{tabular}
\vspace{-5pt}
\caption{Selected LLM models: reasoning and non-reasoning with \textsf{gpt-3.5-turbo} as equivalent to an undergraduate baseline.}
\label{tab:llm_models}
\end{wrapfigure}
Figure~\ref{tab:llm_models} presents a selection of LLMs distinguished by their reasoning capabilities.
Reasoning models like \textsf{o4-mini} and \textsf{o3} excel in complex tasks requiring logical inference, problem-solving, and contextual understanding, such as code debugging, mathematical problem-solving, and decision-making scenarios.
In contrast, non-reasoning models primarily focus on pattern recognition and straightforward tasks like text completion and summarization.
\textsf{gpt-3.5-turbo} serves as a baseline, equating to undergraduate cognitive abilities~\cite{Yeadon2024,Wang2024,LI2025103942}. This model offers a benchmark for deriving actionable insights and recommendations for integrating LLMs into software testing education (refer Section~\ref{rq3_results:sect}).
For convenience, the following notations will be used throughout the paper: \textsf{3.5} for \textsf{gpt-3.5-turbo}, \textsf{4o} for \textsf{gpt-4o}, and \textsf{o4} for \textsf{o4-mini}.

\subsection{Evaluation Metrics}
\label{eval_metrics:sect}
ISTQB exams consist of multiple-choice questions. To evaluate LLM performance, we examine both the selected answers and the accompanying explanations. Each LLM response includes the chosen option, rationale, and reasons for rejecting other choices. These explanations are compared to ISTQB's expected explanations to assess the model's comprehension and decision-making capabilities.

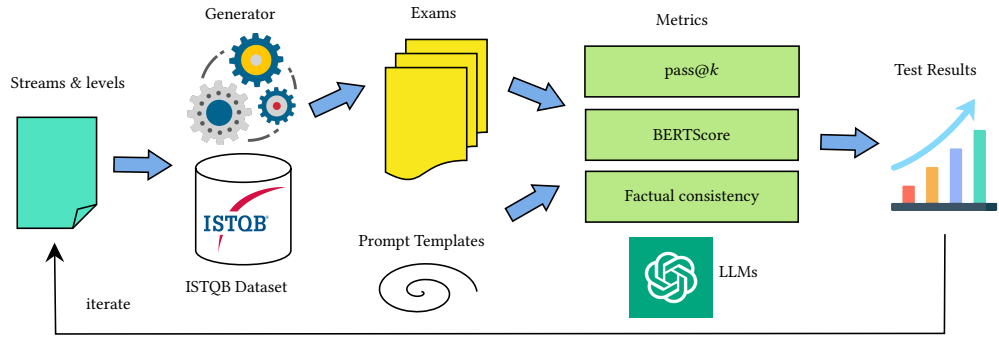
\begin{figure*}[htb]
\centering

\tikzset{every picture/.style={line width=0.75pt}} 
\scalebox{0.8}{
\begin{tikzpicture}[x=0.75pt,y=0.75pt,yscale=-1,xscale=1]

\draw (147.99,76.3) node  {\includegraphics[width=52.5pt,height=52.5pt]{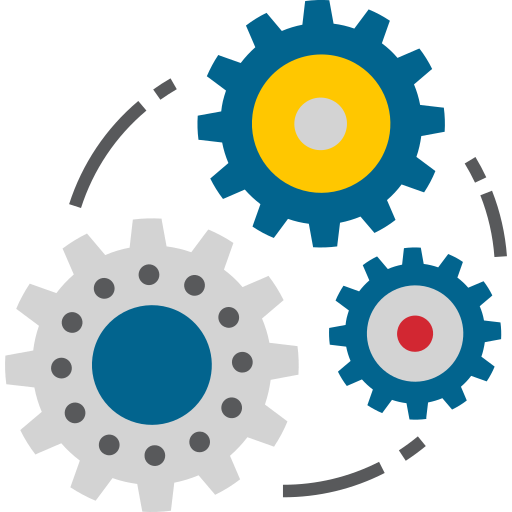}};
\draw   (179,126) -- (179,178) .. controls (179,182.97) and (165.57,187) .. (149,187) .. controls (132.43,187) and (119,182.97) .. (119,178) -- (119,126) .. controls (119,121.03) and (132.43,117) .. (149,117) .. controls (165.57,117) and (179,121.03) .. (179,126) .. controls (179,130.97) and (165.57,135) .. (149,135) .. controls (132.43,135) and (119,130.97) .. (119,126) ;
\draw  [fill={rgb, 255:red, 74; green, 144; blue, 226 }  ,fill opacity=0.74 ] (67.59,119.09) -- (89.28,119.45) -- (89.35,114.85) -- (103.65,124.3) -- (89.05,133.26) -- (89.12,128.66) -- (67.44,128.29) -- cycle ;
\draw  [fill={rgb, 255:red, 248; green, 231; blue, 28 }  ,fill opacity=1 ] (251.47,44.87) -- (302.67,44.87) -- (302.67,103.41) .. controls (270.67,103.41) and (277.07,124.52) .. (251.47,110.86) -- cycle ; \draw  [fill={rgb, 255:red, 248; green, 231; blue, 28 }  ,fill opacity=1 ] (245.07,53.74) -- (296.27,53.74) -- (296.27,112.28) .. controls (264.27,112.28) and (270.67,133.39) .. (245.07,119.73) -- cycle ; \draw  [fill={rgb, 255:red, 248; green, 231; blue, 28 }  ,fill opacity=1 ] (238.67,62.61) -- (289.87,62.61) -- (289.87,121.15) .. controls (257.87,121.15) and (264.27,142.26) .. (238.67,128.6) -- cycle ;

\draw  [fill={rgb, 255:red, 74; green, 144; blue, 226 }  ,fill opacity=0.74 ] (190.59,85.42) -- (210.26,76.3) -- (208.33,72.12) -- (225.31,74.39) -- (216.07,88.82) -- (214.13,84.65) -- (194.46,93.77) -- cycle ;
\draw  [fill={rgb, 255:red, 74; green, 144; blue, 226 }  ,fill opacity=0.74 ] (319.55,68.86) -- (339.71,76.85) -- (341.4,72.57) -- (351.45,86.45) -- (334.62,89.68) -- (336.32,85.4) -- (316.16,77.41) -- cycle ;
\draw  [fill={rgb, 255:red, 74; green, 144; blue, 226 }  ,fill opacity=0.74 ] (314.29,151.53) -- (333.1,140.74) -- (330.81,136.74) -- (347.93,137.53) -- (339.97,152.71) -- (337.68,148.72) -- (318.87,159.51) -- cycle ;
\draw  [fill={rgb, 255:red, 74; green, 144; blue, 226 }  ,fill opacity=0.74 ] (513.31,104.92) -- (534.99,105.49) -- (535.11,100.89) -- (549.32,110.47) -- (534.62,119.29) -- (534.74,114.69) -- (513.07,114.12) -- cycle ;
\draw    (591.33,167.67) -- (591,230.33) -- (30.67,230.67) -- (30.03,176) ;
\draw [shift={(30,173)}, rotate = 89.34] [fill={rgb, 255:red, 0; green, 0; blue, 0 }  ][line width=0.08]  [draw opacity=0] (12.5,-6.01) -- (0,0) -- (12.5,6.01) -- (8.3,0) -- cycle    ;
\draw  [fill={rgb, 255:red, 80; green, 227; blue, 194 }  ,fill opacity=1 ] (41.5,164) -- (6.5,164) -- (6.5,94) -- (56.5,94) -- (56.5,149) -- cycle -- (41.5,164) ; \draw   (56.5,149) -- (44.5,152) -- (41.5,164) ;
\draw   (261.27,204) .. controls (261.8,204) and (262.22,204.14) .. (262.53,204.42) .. controls (262.87,204.71) and (262.92,205.06) .. (262.73,205.44) .. controls (262.5,205.87) and (262.02,206.23) .. (261.27,206.5) .. controls (260.41,206.81) and (259.45,206.94) .. (258.35,206.89) .. controls (257.12,206.83) and (255.99,206.56) .. (254.95,206.08) .. controls (253.8,205.55) and (253,204.86) .. (252.52,204) .. controls (252,203.06) and (251.96,202.09) .. (252.43,201.08) .. controls (252.92,199.99) and (253.92,199.04) .. (255.43,198.23) .. controls (257.06,197.35) and (259,196.78) .. (261.27,196.5) .. controls (263.71,196.21) and (266.12,196.3) .. (268.56,196.78) .. controls (271.14,197.3) and (273.35,198.18) .. (275.16,199.42) .. controls (277.08,200.73) and (278.28,202.26) .. (278.77,204) .. controls (279.28,205.84) and (278.91,207.64) .. (277.69,209.42) .. controls (276.39,211.28) and (274.33,212.84) .. (271.48,214.1) .. controls (268.5,215.42) and (265.09,216.22) .. (261.27,216.5) .. controls (257.27,216.79) and (253.38,216.47) .. (249.6,215.55) .. controls (245.66,214.58) and (242.39,213.09) .. (239.8,211.08) .. controls (237.11,208.99) and (235.51,206.63) .. (235.02,204) .. controls (234.51,201.27) and (235.27,198.63) .. (237.27,196.08) .. controls (239.34,193.45) and (242.49,191.27) .. (246.68,189.57) .. controls (251.02,187.8) and (255.88,186.78) .. (261.27,186.5) .. controls (266.82,186.21) and (272.17,186.75) .. (277.31,188.12) .. controls (282.6,189.54) and (286.93,191.63) .. (290.31,194.42) .. controls (293.79,197.28) and (295.78,200.48) .. (296.27,204) .. controls (296.27,204) and (296.27,204) .. (296.27,204) ;
\draw (149.67,158.67) node  {\includegraphics[width=41pt,height=30.5pt]{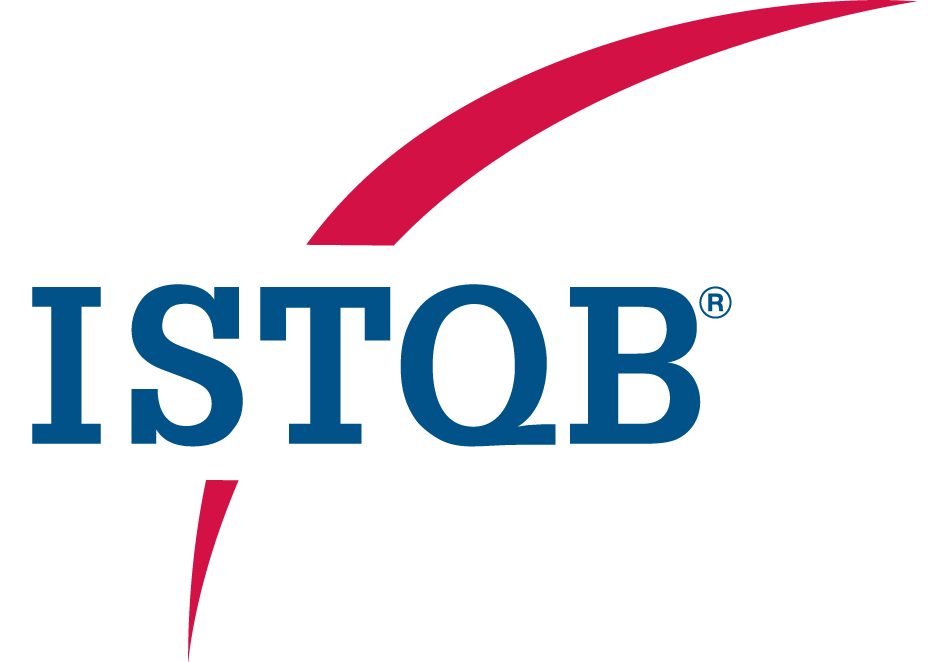}};
\draw (418.57,195) node  {\includegraphics[width=39.25pt,height=37.5pt]{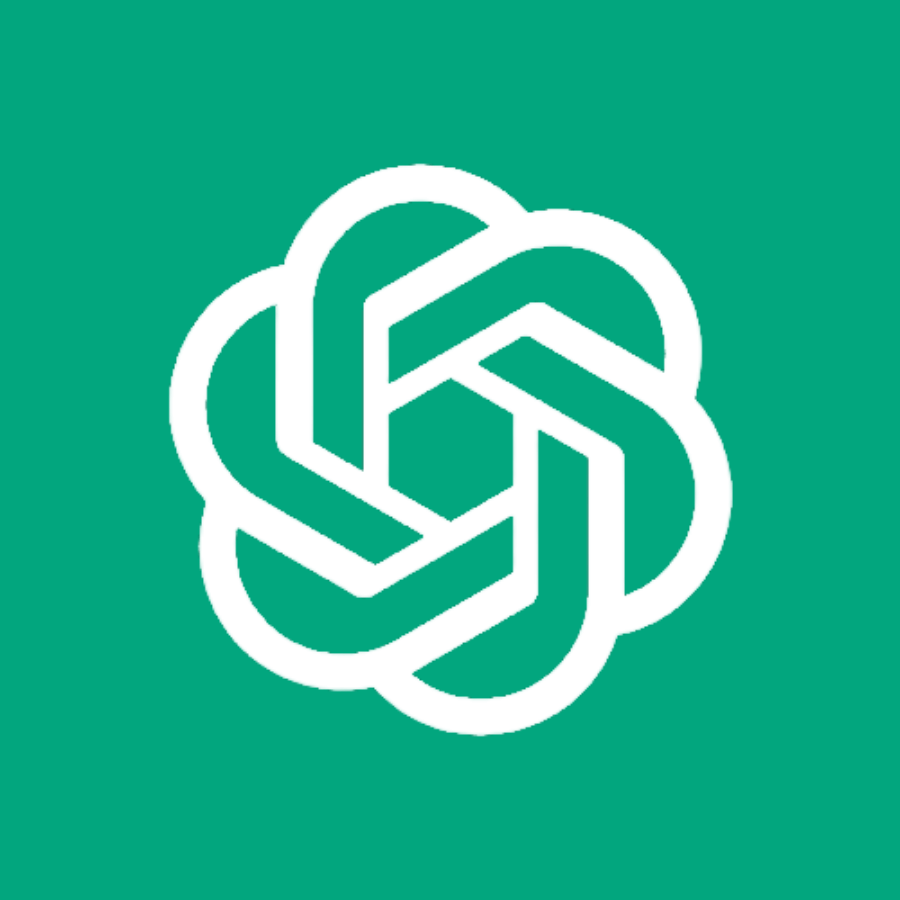}};
\draw (591,118) node  {\includegraphics[width=52.5pt,height=52.5pt]{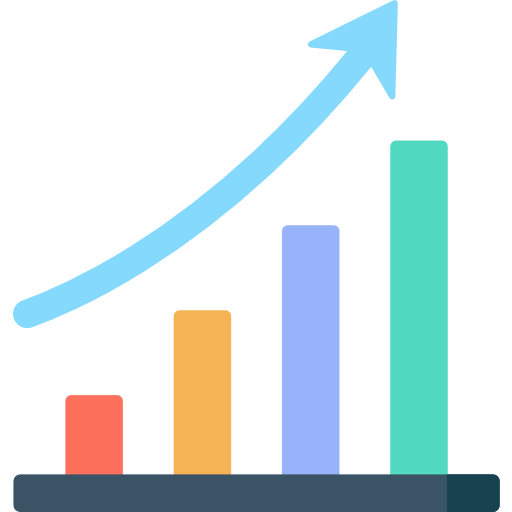}};

\draw (48.53,206.23) node [anchor=north west][inner sep=0.75pt]   [align=left] {{\small iterate}};
\draw (220.27,166.53) node [anchor=north west][inner sep=0.75pt]   [align=left] {{\small Prompt Templates}};
\draw  [fill={rgb, 255:red, 184; green, 233; blue, 134 }  ,fill opacity=1 ]  (364.38,49.08) -- (499.38,49.08) -- (499.38,81.08) -- (364.38,81.08) -- cycle  ;
\draw (431.88,65.08) node   [align=left] {\begin{minipage}[lt]{89.1pt}\setlength\topsep{0pt}
\begin{center}
{\small pass$\displaystyle @k$}
\end{center}

\end{minipage}};
\draw  [fill={rgb, 255:red, 184; green, 233; blue, 134 }  ,fill opacity=1 ]  (364.8,89.08) -- (498.8,89.08) -- (498.8,121.08) -- (364.8,121.08) -- cycle  ;
\draw (431.8,105.08) node   [align=left] {\begin{minipage}[lt]{88.13pt}\setlength\topsep{0pt}
\begin{center}
{\small BERTScore}
\end{center}

\end{minipage}};
\draw  [fill={rgb, 255:red, 184; green, 233; blue, 134 }  ,fill opacity=1 ]  (364.85,128.08) -- (498.85,128.08) -- (498.85,160.08) -- (364.85,160.08) -- cycle  ;
\draw (431.85,144.08) node   [align=left] {\begin{minipage}[lt]{88.6pt}\setlength\topsep{0pt}
\begin{center}
{\small Factual consistency}
\end{center}

\end{minipage}};
\draw (407.87,27.33) node [anchor=north west][inner sep=0.75pt]   [align=left] {{\small Metrics}};
\draw (123.5,23.5) node [anchor=north west][inner sep=0.75pt]  [font=\small] [align=left] {Generator};
\draw (111.33,196) node [anchor=north west][inner sep=0.75pt]   [align=left] {{\small ISTQB Dataset}};
\draw (557.97,58.67) node [anchor=north west][inner sep=0.75pt]   [align=left] {{\small Test Results}};
\draw (447.57,186.3) node [anchor=north west][inner sep=0.75pt]   [align=left] {{\small LLMs}};
\draw (253.24,23.14) node [anchor=north west][inner sep=0.75pt]   [align=left] {{\small Exams}};
\draw (1,66.5) node [anchor=north west][inner sep=0.75pt]   [align=left] {{\small Streams \& levels}};

\end{tikzpicture}
}

\caption{Evaluation framework: Generating random exams from the ISTQB dataset to assess LLM performance.}
\label{eval_framework:fig}
\end{figure*}

We utilize \emph{three} key metrics for evaluation:
\begin{itemize}
    \item \textbf{Functional correctness}~\cite{chen2021codex}: 
    Assessing LLMs by evaluating pass$@k$, where $k$ answers are generated for each problem, and the problem is solved if any answer  matches the official answer key. This metric measures LLM capabilities by accounting for the variability and non-deterministic nature of generative models.
    In the main text, we present pass@$1$ as the  measure of answering accuracy, while more pass@$k$ values are provided in the Appendix for comprehensive evaluation.
    
    \item \textbf{BERTScore}~\cite{Tianyi2020}: Assessing text similarity using contextual embeddings from BERT, allowing a comparison between the LLM's explanation and  expected response based on semantic rather than lexical similarity.
    
    \item \textbf{Factual consistency}~\cite{maynez-etal-2020-faithfulness}: This metric evaluates the accuracy of LLM responses by ensuring alignment with known factual data, which is crucial for distinguishing correct rationales from incorrect ones.
\end{itemize}

These metrics provide a comprehensive assessment of the factual accuracy and reasoning quality of LLM-generated responses, examining their understanding on software testing concepts and effectiveness in complex decision-making scenarios. Since evaluating explanation quality is meaningless for incorrect answers, BERTScore and factual consistency scores are set to zero in such cases.

\subsection{Evaluation Framework}
\label{evaluation_process:sect}


Figure~\ref{eval_framework:fig} illustrates the primary workflow for evaluating LLM performance on ISTQB exams. The iterative process involves several systematic sub-steps.

In the exam generation phase, exams are randomly generated based on stream and level requests, adhering to the ISTQB structure (see Table~\ref{tab:istqb-dataset}). Selection options in each question are also randomly shuffled to ensure validity and reliability in LLM evaluations. 
 \emph{Randomness} is important to 
enhance the validity and reliability of large language model evaluations (see Section~\ref{rq1_results:sect}). 
During the LLM execution sub-step, these exams are combined with prompt templates (see Section~\ref{prompt_engineering:sect}) to create LLM queries.
Finally, the evaluation phase examines LLM responses, using diverse metrics outlined in Section~\ref{eval_metrics:sect}.

\section{Results}
\label{results:sect}

\subsection{RQ1: Evaluating LLMs in  ISTQB Exams}
\label{rq1_results:sect}

\subsubsection{Preliminary Comparison}
\label{surface_comparison}

\begin{table}[tb]
\caption{
LLM performance:
Q-P@1: functional correctness  for question;
E-P@1: functional correctness  for exam;
BE: BERTScore;
FA: factual consistency score.
}

\label{tab:llm_performance}
\small
\centering
\setlength{\tabcolsep}{6pt}
\begin{tabular}{ll|rr|rr}
\rowcolor{gray!25}
\toprule
\textbf{Prompt} & \textbf{LLM}      & \textbf{Q-P@1}  & \textbf{E-P@1}  & \textbf{BE}    & \textbf{FA}\\
\midrule
Basic           & 3.5               &   51.01\%                       &   8.70\%                    &   0.44                   & 0.35\\
Advanced        & 3.5               &   57.14\% ({\color{teal} 6.13$\uparrow$})&   30.43\%         &   0.52                   & 0.46\\
Basic           & 4o                &   69.63\%                       &   69.57\%                   &   0.59                   & 0.45\\
Advanced        & 4o                &   73.97\% ({\color{teal} 4.34$\uparrow$}) &  86.96\%          &   0.67                   & 0.58\\
\hline\hline
Basic           & o4                &   77.67\%                       &   91.30\%                   &   0.68                   & 0.52\\
Advanced        & o4                &   79.26\%  ({\color{teal} 1.59$\uparrow$}) &   91.30\%        &   0.71                   & 0.60\\
Basic           & o3                &   79.79\%                       &   100.00\%                   &   0.70                   & 0.57\\
Advanced        & o3                &   82.75\% ({\color{teal} 2.96$\uparrow$}) &   100.00\%         &   0.75                   & 0.68\\
\bottomrule
\end{tabular}
\end{table}

Table~\ref{tab:llm_performance} summarizes the performance of four LLMs on 23 exam samples, each corresponding to an ISTQB exam type listed in Table~\ref{tab:istqb-dataset}. For each LLM and prompt combination, average values across multiple evaluation metrics are reported.
Detailed performance  of LLMs and prompts and their costs are given in the Appendix.

Due to the limited sample size of one per most test type as provided by ISTQB, we characterize this experiment as a preliminary comparison. Our objective is to maintain the authenticity of the ISTQB dataset while rigorously evaluating the actual capabilities of LLMs and avoiding potential biases introduced by synthetic data.
Subsequently, Section~\ref{deep_comparison} presents a deep comparison utilizing multiple authentic test samples from the ISTQB CTFL.

\begin{table*}[tb]
\caption{
Deep comparison for LLM performance on CTFL v4.0.1:
Q-P@1: functional correctness  for question running;
E-P@1: functional correctness  for exam running;
BE: BERTScore;
FA: factual consistency score.
‘/20’ indicates averages over 20 test runs in this comparison; ‘/1’ indicates the corresponding single test run in Table~\ref{tab:llm_performance}.
}
\label{tab:llm_performance_ctfl4.0_deep}
\small
\centering
\setlength{\tabcolsep}{6pt}
\begin{tabular}{ll|rrrr|rrrr}
\rowcolor{gray!25}
\toprule
\textbf{Prompt} & \textbf{LLM}   & \textbf{Q-P@1/20} & \textbf{Q-P@1/1} & \textbf{E-P@1/20} & \textbf{E-P@1/1} & \textbf{BE/20} & \textbf{BE/1} & \textbf{FA/20} & \textbf{FA/1}\\
\midrule
Basic           & 3.5      &   43.25\%  &  55.00\%        &   0.00\%     & 0.00\%           &   0.38 &  0.49           & 0.29 &  0.31\\
Advanced        & 3.5      &   47.50\%    &  60.00\%      &   0.00\%     & 0.00\%           &   0.44 &  0.55           & 0.40 &  0.52\\
Basic           & 4o       &   71.25\%  &  77.50\%        &   90.00\%    &  100.00\%         &   0.62 &  0.64           & 0.51 &  0.57\\
Advanced        & 4o       &   75.63\%  &  77.50\%        &   95.00\%    &  100.00\%         &  0.70 &  0.72           & 0.64 &  0.66\\
\hline\hline
Basic           & o4       &   85.50\%  &  87.50\%        &   100.00\%   &  100.00\%         &   0.76 &  0.78           & 0.58 &  0.63\\
Advanced        & o4       &   88.75\%  &  90.00\%        &   100.00\%   &  100.00\%         &   0.82 &  0.83           & 0.65 &  0.77\\
Basic           & o3       &   89.50\%  &  95.00\%        &   100.00\%   &  100.00\%         &   0.81 &  0.86           & 0.66 &  0.77\\
Advanced        & o3       &   89.50\%  &  95.00\%        &   100.00\%   &  100.00\%         &   0.83 &  0.88           & 0.73 &  0.81\\
\bottomrule
\end{tabular}
\end{table*}

The results show that the advanced prompt consistently yields higher question accuracy across all evaluated models, with a notable improvement of 6.13\% observed for \textsf{gpt-3.5}. Interestingly, despite both the basic and advanced prompts for \textsf{gpt-3.5} yielding question accuracies 
approximately centered around 54\%,
the advanced prompt exhibited a substantially higher exam pass rate, increasing by 21.73\%. 

As anticipated, models equipped with enhanced reasoning capabilities outperform non-reasoning models in terms of correctly answered questions. Specifically, \textsf{o3} achieves the highest pass rates, reaching 100\%, whereas \textsf{gpt-3.5} exhibits the lowest pass rate of 8.7\% when using the basic prompt.
Among the reasoning models, \textsf{o3} attains higher accuracy compared to \textsf{o4}, indicating better reasoning performance.
Although the intrinsic capabilities of LLM architectures significantly influence performance, our findings highlight the critical role of advanced prompt engineering in enhancing model outcomes.

Moreover, the two additional evaluation metrics---BERTScore (BE) and factual consistency (FA)---exhibit strong agreement. Models achieving higher BE scores tend to also attain higher FA scores, indicating a close relationship between semantic similarity and factual consistency in model outputs.
Notably, the \textsf{o3} model, when paired with the advanced prompt, demonstrates the best performance, achieving an average question accuracy of 82.75\%. Its explanation quality is supported by 0.75 BE score, indicating good semantic alignment, and 0.68 FA score, reflecting moderate factual reliability~\cite{LinkedinPost2024}.

The modest accuracy exhibited by the baseline model \textsf{gpt-3.5} underscores the complexity of our dataset and evaluation framework. Using authentic original exam samples, our results confirm that introducing randomness through the shuffling of selection options enhances the validity and reliability of LLM evaluations. 

\subsubsection{Deep Comparison}
\label{deep_comparison} 
Table~\ref{tab:llm_performance_ctfl4.0_deep} presents a comprehensive comparison based on 20 randomly generated exam samples (each with 40 questions) drawn from a pool of 160 authentic ISTQB CTFL v4.0.1 questions, strictly adhering to the ISTQB exam format. We report average metric values across these samples to evaluate model performance.
A similar experiment can be done for CTFL v3.1 but not for other ISTQB tests due to the limitation of authentic questions.

The reduction in Q-P@1/20 and E-P@1/20 accuracy observed in this deep comparison, compared to Q-P@1/1 and E-P@1/1 of the corresponding single test run in the preliminary comparison, reflects a higher level of difficulty inherent in the deep experimental setup. This increased challenge stems from the introduction of additional randomness in the deep experiment, not only from the order of selection options (as in the preliminary comparison) but also from variations in the combination set of questions (refer to Section~\ref{evaluation_process:sect}). This evidence further corroborates the reliability and stability of our evaluation framework and results.

\subsection{RQ2: Optimising Prompt Engineering to Enhance LLM Responses}
\label{rq2_results:sect}
Tables~\ref{tab:llm_performance}-\ref{tab:llm_performance_ctfl4.0_deep} illustrate the effectiveness of the advanced prompt in enhancing the performance of LLMs. Developing such a prompt presents a non-trivial challenge, as it must be self-contained, user-friendly for non-expert users (see Section~\ref{prompt_engineering:sect}), and robust across multiple LLM architectures when applied to diverse ISTQB exam types. This contribution underscores the critical role of prompt engineering in optimizing LLM outputs within the domain of software testing certification.
\smallskip
\begin{myprompt}
\textbf{Advanced.} 
Act as a \emph{certified ISTQB \{exam acronym \} trainer}. For the following multiple-choice question, provide:
\begin{itemize}
    \item The correct options(s).
    \item An explanation of why the option(s) is correct.
    \item An explanation of why the other option(s) are correct.
    \item Ensure all parts align with the \emph{official ISTQB \{exam acronym\} syllabus and principles}.
\end{itemize}
\{ISTQB question\}
\end{myprompt}

\subsection{RQ3: Harnessing LLMs for Learning and Teaching in Software Testing}
\label{rq3_results:sect}

\subsubsection{For Educators}
\label{rq3_educator:sect}

Figure~\ref{heat_map:passing:k:core:fig} presents the answering accuracy of the baseline model, \textsf{gpt-3.5}, with the basic prompt across different cognitive levels, as defined in Table~\ref{tab:istqb-dataset}.
The results can be further detailed by providing statistical breakdowns of the learning chapters within each cognitive level.
Due to space limitation, only the results for core-level exams are presented here, with the full results given in the Appendix.

\begin{figure}[tb]
\centering
\setlength{\tabcolsep}{6pt}
\begin{tikzpicture}[scale=0.6]
  \foreach \y [count=\n] in {
      {67,52,50,---},
      {83,56,43,---},
      {---,58,30,57},
      {---,50,14,17},
      {---,47,50,29},
      {---,65,29,17},
      {---,45,59,50},
    } {
      \ifnum\n<5
        \node[minimum size=6mm] at (\n, 0) {K\n};
      \fi
      \foreach \x [count=\m] in \y {
        \node[fill=yellow!\x!purple, minimum size=6mm, text=white] at (\m,-\n) {\x};
      }
    }

  \foreach \a [count=\i] in {CTFL v4.0.1,CTFL v3.1,CTAL-TM v3.0,CTAL-TA v4.0,CTAL-TA v3.1,CTAL-TAE v2.0,CTAL-TTA v4.0} {
    \node[minimum size=6mm] at (-1.5,-\i) {\a};
  }
\end{tikzpicture}
\caption{Q-P@1 across cognitive levels for  core-level exams using the baseline model and basic prompt.  '---' means the corresponding level is not applicable.}
\label{heat_map:passing:k:core:fig}
\end{figure}
The baseline model with the basic prompt achieves an E-P@1 accuracy of 8.7\%, highlighting the inherent difficulty of ISTQB exams for students.
This model is designed to approximate the cognitive performance level of undergraduate students~\cite{Yeadon2024, Wang2024, LI2025103942}.
For example, in the CTAL-TM v3.0, the lowest accuracy for Q-P@1 was observed at the K3 level across the evaluated levels K2 to K4. This finding aligns with our  experience in ISTQB preparation courses, where students might find some lower-level questions more difficult than those classified at higher cognitive levels.

The baseline model's performance offers educators valuable insights into the challenging aspects of the ISTQB syllabus for undergraduates. 
This aids in creating instructional materials and effective teaching strategies, aimed at enhancing student understanding across various learning areas.

\subsubsection{For Students}
\label{rq3_student:sect}

Integrating LLMs into ISTQB-based software testing education offers multiple advantages, including the role of a virtual tutor that provides students with real-time feedback and fosters self-directed learning.
It is important to emphasize that while the virtual tutor delivers valuable support, it is not intended to replace human educators.

The \textsf{o3} model combined with the advanced prompt exhibits strong potential as a virtual tutor. It achieves a high question accuracy of 82.75\%, accompanied by good semantic alignment and moderate factual reliability in explanation quality, as reflected by BE and FA scores of 0.75 and 0.68, respectively (see Table~\ref{tab:llm_performance}). Moreover, the model attains a 100\% exam pass rate.

\section{Conclusions and Future Work}
\label{conclusions:sect}

This research investigates LLM performance using a comprehensive ISTQB dataset, providing actionable insights and recommendations for integrating LLMs into software testing education based on ISTQB curricula.

Despite the promising outcomes of our research, several limitations remain. Our evaluation does not incorporate time constraints that could impact performance, typically present in ISTQB exams. Furthermore, the study does not explore all leading LLMs, such as Gemini, Claude, or Llama, nor does it evaluate LLM capabilities at the Expert level, as noted in Section~\ref{istqb_benchmark:sect}. The complexity of Expert essay questions requires an advanced evaluation mechanism. Additionally, techniques like few-shot prompts, chain-of-thought reasoning, and RAG are impractical for typical students and educators due to their complexity. Future research should consider these limitations to refine LLM evaluation methodologies further.


\bibliographystyle{ACM-Reference-Format}
\bibliography{references}


\end{document}